\documentclass[preprint]{rspublic}

\usepackage{amsbsy}
\usepackage{amssymb}
\usepackage{amsmath}
\usepackage{natbib}

\begin{document}
\title[Variational heat conductivity]{Variational multi-fluid dynamics and causal heat conductivity}

\author{N. Andersson$^1$ and G. L. Comer$^2$}
\affiliation{$^1$ School of Mathematics, University of Southampton, UK \\
$^2$ Department of Physics and Center for Fluids at All Scales, Saint Louis University, USA}

\maketitle

\def\be{\begin{equation}}
\def\ee{\end{equation}}
\def\n{{\rm n}}
\def\s{{\rm s}}
\def\N{{\rm N}}
\def\S{{\rm S}}
\def\mun{{\mu_\n}}
\def\mus{{\mu_\s}}

\begin{abstract}{Fluid dynamics; Heat conductivity}
We discuss heat conductivity from the point of view of a variational
multi-fluid model, treating entropy as a dynamical entity.
We demonstrate that a two-fluid model with a
massive fluid component and a massless entropy can reproduce a
number of key results from extended irreversible thermodynamics.
In particular, we show that the entropy entrainment is intimately
linked to the thermal relaxation time that is required to make heat propagation in solids causal. We also discuss non-local terms that arise naturally in a dissipative
multi-fluid model, and relate these terms to those of
phonon hydrodynamics. Finally, we formulate a complete heat conducting
two-component model and discuss briefly the new dissipative terms that arise.
\end{abstract}

\section{Introduction}

Heat conductivity is a central problem in thermodynamics. It is well known that the classical description
in irreversible thermodynamics, essentially Fourier's law, has  unattractive features. In particular,
it predicts an instantaneous propagation of thermal signals. This is in contradiction to the expected hyperbolic
nature of physical laws. In fact, the associated non-causality would be completely unacceptable within a
relativistic theory. Resolving this issue has been a main motivating factor behind the development of
extended irreversible thermodynamics \citep{joubook,mullerbook}, a model which introduces additional dynamical fields
in order to retain hyperbolicity and causality. This model has proved useful in various application areas,
ranging from superfluid systems to heat conduction in solids.

In this paper we consider the problem of heat conductivity within the flux-conservative multi-fluid framework developed by \citet{monster} (with the
corrections discussed by \citet{erratum}). This model builds on the non-dissipative variational model developed by \citet{prix}, and represents
the natural non-relativistic counterpart to Carter's convective variational hydrodynamics in general relativity  \citep{carter}, see
\citet{livrev} for a recent review. These models have so far primarily been used to investigate the dynamics of compact stars (see \citet{ac01,prec1,glitch,andrea} for details).
Yet, one would expect the general framework to be \underline{universally} relevant. It is therefore of some interest to consider  applications in other
problem areas. This serves several purposes. First of all, it is often the case that the elegant geometric view of a relativistic analysis (where time and space are treated ``equally'') simplifies
the description of a complex system. The close link between the variational multi-fluid model and
the relativistic counterpart may lead to insights that would be hard to reach otherwise. Secondly,
it is interesting to learn from the requirements of models for a range of different systems. Ultimately, this will improve our understanding
of the general multi-fluid framework, and the role of the various parameters in the model.

In this paper we focus on the simplest ``conducting'' system, with a single species of particle together with a massless
entropy component. If the entropy drifts relative to the particles, the system is heat conducting. Throughout
the discussion we will assume that the entropy can be treated as a (massless) fluid. In essence, this means that the
``phonon mean free path'' is not too large.
We aim to demonstrate that our flux-conservative formulation for this system captures key aspects of extended thermodynamics.
We illustrate this by writing down a model for heat conductivity in a rigid solid and comparing to various results in the literature.
This exercise  makes it clear that our model incorporates a finite propagation speed for heat. We also learn that the
associated relaxation timescale is directly linked to the entrainment between  particles and  entropy.
This entropy entrainment also played a central role in our recent discussion of finite temperature superfluids
\citep{helium}.
In fact, the model we consider here is formally equivalent to our recent model for superfluid Helium, the key element being that the massless
entropy component is allowed to flow relative to the particles in the system. However, the discussion here differs in that we do not impose the irrotationality
constraint associated with a superfluid system. Instead, we focus on the heat conductivity.
We wish to understand to what extent this, conceptually rather elegant, model captures the complex
physics associated with heat flow. As an interesting by-product, the
present discussion suggests how the Helium model
could be extended (using a three-fluid model) to account for the thermal conductivity associated with the interaction between rotons and phonons
that dominates at higher temperatures \citep{khalatnikov}. This is an interesting problem since the development of a causal model for this phenomenon is
still outstanding. Our analysis also hints at the key ingredients of a causal relativistic model for heat conductivity. The main lesson
is that the entropy entrainment must be retained in order to avoid pathological behaviour.

\section{Multi-Fluid formulation}

We consider a simple system with two dynamical degrees of freedom. We distinguish between
the mass carrying ``atoms'' and the massless ``entropy''. The former will be identified by constituent index $\n$, it has mass $m$, number density $n$ and flows with a velocity
$v_\n^i$, while the latter is represented by $\s$, with number density $s$ (the entropy per unit volume)
and a flow given by $v_\s^i$.
The canonical momentum (density) for each fluid is determined by the variational analysis of \cite{prix}. This leads to
\be
\pi_i^\n = mn v_i^\n - 2 \alpha w_i^{\n\s} \ ,
\ee
for the atoms and
\be
\pi^\s_i = 2 \alpha w_i^{\n\s} \ ,
\label{smom}\ee
for the entropy. This latter relation encodes the inertia of heat. We see that the heat's inertia vanishes
if the entrainment between the two components, quantified by $\alpha$, vanishes.
In these expressions we have used the relative velocity $w_i^{\n\s} = v_i^\n-v_i^\s$. We recall that the
variational analysis takes as its starting point an energy functional $E$, representing the
equation of state. The coefficient $\alpha$ then follows from
\be
\alpha = \left. {\partial E \over \partial w_{\n\s}^2} \right|_{n,s} \ .
\ee

As discussed by \cite{monster,helium} the momentum equations for the dissipative two-component
system can be written
\be
f_i^\n = \partial_t \pi_i^\n + \nabla_j (v_\n^j \pi_i^\n + D^{\n j}_{\ \ i} ) + n \nabla_i \left( \mu_\n
- { 1 \over 2} m v_\n^2 \right) + \pi_j^\n \nabla_i v_\n^j \ ,
\label{eulern}\ee
and
\be
f_i^\s = \partial_t \pi_i^\s + \nabla_j (v_\s^j \pi_i^\s + D^{\s j}_{\ \ i} ) + s \nabla_i T + \pi_j^\s \nabla_i v_\s^j \ .
\label{eulers}\ee
Here the terms $D^\n_{i j}$ and $D^\s_{i j}$ represent the dissipation and $f^i_\n$ and $f^i_\s$ are the respective forces acting on the matter and entropy.
The chemical potentials are determined by
\be
\mu_\n = \left. {\partial E \over \partial n} \right|_{s, w_{\n\s}^2} \ ,
\ee
and
\be
\mu_\s = \left. {\partial E \over \partial s} \right|_{n, w_{\n\s}^2} \equiv T \ ,
\label{tdef}\ee
where we have identified the temperature, $T$, in the usual way (note that we use units such that Boltzmann's constant is equal to unity, $k_B=1$)\footnote{It should be pointed out that the
notion of temperature in a non-equilibrium system is non-trivial, see \citet{cvj} for a thorough discussion. In the present analysis we assume that the temperature is obtained from \eqref{tdef}, i.e. in the same way as in thermal equilibrium. This ``operational'' definition seems the most natural in the present context.  }.

Since we have no particle creation or destruction, mass conservation leads to
\be
\partial_t n + \nabla_j (n v_\n^j ) = 0 \ .
\label{conn}\ee
At the same time the entropy can increase, so we have
\be
\partial_t s + \nabla_j (s v_\s^j ) = \Gamma_\s \ ,
\label{cons}\ee
where the second law of thermodynamics requires $\Gamma_\s \ge 0$.
Finally, the dissipative terms are constrained by the fact that we
consider a closed system. As discussed by \citet{monster}, this means that
we must have
\be
f_i^\s = - f_i^\n \ ,
\ee
and
\be
D_{ij}^\s = D_{ij} - D_{ij}^\n \ .
\ee
Here, $D_{i j}$ is the total ``dissipation''. It
is distinguished by the fact that it is symmetric
in its indices whereas $D_{ij}^\n$ and $D_{ij}^\s$ do not have to be.

Following the steps taken in \cite{helium}, i.e. constructing the dissipative terms
from the relevant thermodynamic fluxes and taking account of the
Onsager symmetry, we arrive at\footnote{We use a coordinate basis to represent tensorial relations.
In other words, we distinguish between co- and contra-variant objects, $v_i$ and $v^i$, respectively.
Indices, which range from 1 to 3, can be raised and lowered with the (flat space) metric $g_{ij}$, i.e.,
$v_i = g_{ij} v^j$. Derivatives are expressed in terms of the covariant derivative
$\nabla_i$ which is consistent with the metric in the sense that $\nabla_i g_{kl} = 0$.
This formulation has great advantage when one wants to discuss the geometric nature of the
different dissipation coefficients. We also use the volume form $\epsilon_{ijk}$ which is completely antisymmetric,
and which has only one independent component (equal to $\sqrt{g}$ in the present case).}
\be
- f_i^\n = 2 \mathcal{R}^{\n\n} w^{\n\s}_i + 2 \mathcal{S}^{\n\n} W^{\n\s}_i \ ,
\label{f1}\ee
\be
- D_{ij}^\n = \mathcal{S}^{\n\n} \epsilon_{ijk} w_{\n\s}^k + g_{ij} (\zeta^{\n\n} \Theta_{\n\s} + \zeta^\n \Theta_\s)
+ 2 \eta^{\n\n} \Theta^{\n\s}_{ij} + 2 \eta^\n \Theta^\s_{ij} + \sigma^{\n\n} \epsilon_{ijk} W_{\n\s}^k \ ,
\label{f2}\ee
and
\be
- D_{ij} =  g_{ij} (\zeta^{\n} \Theta_{\n\s} + \zeta \Theta_\s)
+ 2 \eta^{\n} \Theta^{\n\s}_{ij} + 2 \eta \Theta^\s_{ij} \ .
\label{f3}\ee
Here we have defined the expansion
\be
\Theta_\s = \nabla_j v_\s^j \ ,
\ee
the trace-free shear
\be
\Theta^\s_{ij} = { 1 \over 2} \left( \nabla_i v^\s_j + \nabla_j v^\s_i - { 2 \over 3} g_{ij} \Theta_\s \right) \ ,
\ee
and the ``vorticity''
\be
W_\s^i = { 1 \over 4} \epsilon^{ijk}( \nabla_j v^\s_k - \nabla_k v^\s_j) \ ,
\ee
of the entropy flow.
The quantities
$\Theta_{\n\s}$, $\Theta^{\n\s}_{ij}$  and $W_{\n\s}^k$ are constructed from $w_{\n\s}^i$ analogously.

So far, the development has been  formal.
The model was developed by combining the non-dissipative equations of motion from the variational analysis \citep{prix}
with the general form for the dissipative terms assuming a quadratic deviation from thermodynamic equilibrium. We have accounted for the
Onsager symmetry between the dissipative terms, and  imposed the conservation of total angular momentum. This reduces the problem to one with nine, as yet unspecified, dissipation coefficients;
$\mathcal{R}^{\n\n}$, $\mathcal{S}^{\n\n}$,  $\zeta^\n$,  $\eta^\n$,  $\zeta^{\n\n}$, $\eta^{\n\n}$, $\sigma^{\n\n}$, $\zeta$ and $\eta$.

It should be noted that a number of, essentially, analogous formulations for
multi-component systems exist in the literature, see for example \citet{drew,raja}. However, most
of these are not arrived at by variational arguments, and hence they do not distinguish the fluxes
from the canonically conjugate momenta. This means that the entrainment effect tends to be ignored.
In fact, the novelty of our approach relates to the entrainment between particles and entropy.
From an intuitive point of view, this encodes the inertia of heat and allows us to assign an
effective mass to the entropy component. It is well known that these concepts are central to
causal models of heat conductivity, and we will soon see why this is so.

\section{The standard approach to heat conductivity}

It is natural to begin the discussion of heat conductivity by sketching the classic approach to the problem.
This provides a useful contrast to the more general model that we will develop.

The following argument is more or less identical to that  provided by \citet{prix}. First we rewrite \eqref{cons} as
\be
\partial_t s + \nabla_i [ s v_\n^i + s w_{\s\n}^i] = \Gamma_\s \ .
\ee
In general, $\Gamma_\s > 0$ if the system is out of equilibrium.
Next we introduce the heat flux vector
\be
q^i = sT w_{\s\n}^i \ ,
\label{qdef}\ee
such that
\be
\partial_t s + \nabla_i \left[ s v_\n^i + {q^i \over T} \right] = \Gamma_\s \ .
\ee
Keeping only the linear friction term associated with $ \mathcal{R}^{\n\n}$, and using eq.~(2.9) from \citet{helium}, i.e.
\be
   T \Gamma_\s = - f_i^\n w_{\n\s}^i - D^j_{\ i} \nabla_j v_\s^i -
   D^{\n j}_{\ i} \nabla_j w_{\n\s}^i \ ,
   \label{TGs}
\ee
we arrive at
\be
\Gamma_\s = - {1 \over T} f_i^\n w_{\n\s}^i =   {2 \over T} \mathcal{R}^{\n\n} w_{\n\s}^2 =  2\mathcal{R}^{\n\n} \left( {  q \over s T} \right)^2  \ge 0 \ .
\ee
The second law of thermodynamics thus requires that $ \mathcal{R}^{\n\n} \ge 0$.
In the following we will often omit the subscripts on the resistivity coefficient $\mathcal{R}^{\n\n}$. This makes the equations
somewhat clearer and we do not believe that it should cause any confusion.

Finally, if we consider the case of vanishing entrainment (cf. \citet{prix}) then it follows from \eqref{eulers} that
\be
f_i^\s = s \nabla_i T = - f_i^\n = 2 \mathcal{R} w_i^{\n\s}  = - 2 { \mathcal{R} \over sT} q_i \ ,
\ee
or
\be
q_i = - { 1 \over 2} { s^2 T \over \mathcal{R} } \nabla_i T \ .
\ee
Comparing to  Fourier's law we  identify the thermal conductivity as
\be
\kappa = { 1 \over 2} { s^2 T \over \mathcal{R} } \ .
\ee
This completes the traditional description of the heat conductivity problem. However, for reasons that we have already discussed,
this analysis is not entirely satisfactory.
Fortunately, the multi-fluid approach advocated here allows for a more general view. This alternative description has a richer phenomenology
and recovers a number of features of extended irreversible thermodynamics \citep{joubook,mullerbook}.
Illustrating this is one of the main purposes of our discussion.

\section{The extended thermodynamics view}

The analysis in the previous section focussed on the entropy conservation law.
The spirit of the discussion was very much that of classic irreversible thermodynamics.
We will now depart from this view, and ask what we can learn by taking the
two-fluid model, e.g. the entropy inertia, at face value. The massless entropy flow and the associated entrainment then play
central roles.

Let us consider the entropy momentum equation (\ref{eulers}) a bit closer. We can rewrite
this  equation as
\begin{multline}
(\partial_t + v_\n^j \nabla_j ) \pi_i^\s - \nabla_j \left( {1 \over 2 \alpha} \pi^j_\s \pi^\s_i \right)
- \pi_j^\s \nabla_i \left( {1 \over 2 \alpha} \pi^j_\s \right) \\
+ s \nabla_i T + \pi_i^\s (\nabla_j v_\n^j)
+ \pi^\s_j\nabla_i v_\n^j + \nabla_j D^{\s j}_{\ \ i} = f_i^\s \ .
\label{masters}\end{multline}
For simplicity, we first include only the resistive contribution to the dissipation, i.e., we set
\be
D^\s_{ij} = 0 \ , \qquad  \mbox{ and } \qquad f^\s_i = 2\mathcal{R} w_i^{\n\s} \ ,
\ee
as before.
Then we have
\begin{multline}
(\partial_t + v_\n^j \nabla_j ) \pi_i^\s - \nabla_j \left( {1 \over 2 \alpha} \pi^j_\s \pi^\s_i \right)
- \pi_j^\s \nabla_i \left( {1 \over 2 \alpha} \pi^j_\s \right) \\
+ s \nabla_i T + \pi_i^\s (\nabla_j v_\n^j)
+ \pi^\s_j\nabla_i v_\n^j =  { \mathcal{R} \over \alpha } \pi_i^\s \ ,
\end{multline}
where we recall the definition of the entropy momentum, \eqref{smom}.

It turns out to be instructive to view this equation in the context of extended irreversible thermodynamics \citep{joubook,mullerbook}.
To facilitate the comparison, let us focus on the simplified case of  heat conduction in a rigid solid. In that case we
have $v_\n^i = 0$ and our momentum
equation simplifies to
\be
\partial_t \pi_i^\s - \nabla_j \left( {1 \over 2 \alpha} \pi^j_\s \pi^\s_i \right)
- \pi_j^\s \nabla_i \left( {1 \over 2 \alpha} \pi^j_\s \right) + s \nabla_i T  =  { \mathcal{R} \over \alpha } \pi_i^\s \ ,
\ee
Moreover, as before it makes sense to introduce the heat flux $q^i$ as
\be
w_{\s\n}^i = {1 \over sT} q^i = - { 1 \over 2 \alpha} \pi^\s_i \ .
\label{wdef}\ee
In the following we will often choose to keep the relative velocity as the main variable (we
also omit the constituent indices, using $w^i = w_{\s\n}^i$ for clarity). This simplifies many of the equations as
it suppresses factors of $sT$. Expressing all results in terms of the heat flux vector is trivial
given \eqref{wdef}.

It follows that
\be
- \partial_t \left({ 2 \alpha w_i} \right) - \nabla_j \left( 2 \alpha w^j w_i  \right)
-  \alpha   \nabla_i w^2 + s \nabla_i T  +    2 \mathcal{R} w_i
 = 0 \ .
\label{cat1}\ee
This is one of our main results.

Consider the limit of vanishing entrainment. Letting $\alpha \to 0$  we are left with
\be
w_i = - { s \over 2 \mathcal{R} } \nabla_i T \ ,
\ee
or
\be
q_i = - { s^2 T \over 2 \mathcal{R} } \nabla_i T = - \kappa \nabla_i T \ ,
\ee
where we have identified the thermal conductivity, $\kappa$, as in the previous section. In the limit
of vanishing entrainment we
recover Fourier's law, as expected.

If we instead linearise
the equation (with respect to thermal equilibrium, $q^i = 0 $)  we
find that
\be
- \partial_t \left( {2 \alpha \over sT} q_i \right)  + s \nabla_i T  +   { 2  \mathcal{R} \over sT} q_i
 = 0 \ .
\label{linear}\ee
For constant parameters\footnote{There is no physical reason why $\alpha/sT$ should be "constant". We only make the assumption in order to facilitate a direct comparison with the Cattaneo equation.
The discussion of \citet{morro} makes it quite clear that the parameter needs to be temperature dependent in order
for this kind of model to be able to reproduce data from second sound experiments.}, this can be written
\be
\tau \partial_t q_i + q_i = - \kappa \nabla_i T \ ,
\label{cattaneo}\ee
where we have introduced the thermal relaxation time, $\tau$, according to
\be
\tau = - { \alpha \over \mathcal{R}} \ .
\label{tau}\ee
Eq. (\ref{cattaneo}) is known as the Cattaneo equation \citep{cattaneo}. It resolves
the ``paradox'' associated with instantaneous propagation of heat predicted in classic irreversible
thermodynamics, see \citet{jcv} for a brief discussion.
It also leads to the presence of a second sound in solids,
an effect that has been observed in  laboratory experiments on dielectric crystals, see for example the discussion by
\citet{rms}. As described by \citet{joubook}, the
 Cattaneo equation provided key stimulus for
the development of extended irreversible thermodynamics.

The above exercise shows that the variational
multi-fluid formalism contains a key element of extended thermodynamics.
Moreover, we have a natural interpretation of the entropy entrainment, $\alpha$. In terms of the
relaxation time $\tau$ we have
\be
\alpha = - \tau \mathcal{R} = - { s^2 T \over 2 } \left( {\tau \over \kappa} \right) \ .
\ee
That is, we should expect to have $\alpha \le 0$.

Is it surprising that the Cattaneo equation can be deduced from the variational two-fluid model?
Not really. In retrospect the result is, more or less, obvious.
This is easy to see if we compare the nature of the variational energy functional $E$ \citep{prix}
to the generalised entropy used in extended irreversible thermodynamics \citep{jcv}.
In the variational case, we have $E=E(n,s,w^2)$ and in the case of a solid (taking $n$ to be constant)
it follows that
\be
dE = {\partial E \over \partial s} ds + {\partial E \over \partial w^2 } dw^2 = T ds + \alpha dw^2 =
T ds + \alpha d \left({q \over sT}\right)^2 \ .
\label{gibbs1}\ee
Meanwhile, one of the tenets of extended irreversible thermodynamics is the generalised entropy
\be
s(u,q)=s_0(u)-{ 1 \over 2} \beta(T) q^2 \ ,
\ee
where $u$ is the internal energy. As discussed by \citet{alvarez}, this assumption is common to a number of alternative
approaches to irreversible thermodynamics. The entropy satisfies the generalised Gibbs identity
\be
ds = { 1 \over T} du - \beta dq^2 \ .
\label{gibbs2}\ee
It is now apparent that, once we identify $u=E$, \eqref{gibbs1} and \eqref{gibbs2} contain
similar information. The key point is that the energy/entropy of the system depends  on the
heat flux. This is natural since the heat flux corresponds to the flow of energy relative to the matter
current. The variational approach  provides a slightly different perspective on the problem, but the
predicted dynamics should be equivalent.

Before moving on, it is worth discussing the role of the
non-linear terms that were discarded in \eqref{linear}.
The variational analysis naturally leads to the presence of quadratic terms in the
heat flux. Due to their origin, i.e. the entropy momentum equation, these terms (essentially) take the
same form as the non-linear terms in the standard Euler equation.
In other words, they do not represent the most general non-linearities that one might envisage (in the dissipative
problem).
In the interest of clarity, we have chosen not to compare the non-linear terms in our model to
various attempts at constructing non-linear heat-conducting models, see for example
\citet{mr87,rms,jlmp,lrv} and \citet{llebot}. Such a comparison would obviously be interesting
given that non-linearities are  relevant for the development of both shocks and turbulence.
However, our main initial aim is to establish the viability of the
multi-fluids approach to the heat problem. For this purpose, the evidence provided by the linear comparison should be adequate.

\section{Including ``non-local'' terms}

The model discussed in the previous section is the simplest in a hierarchy of possible
heat conductivity models. By relaxing the assumptions that led to \eqref{cattaneo} we can easily obtain
more complicated models. Such models are interesting, because \eqref{cattaneo} may not
provide a faithful representation of all relevant phenomena. It is
obviously important to compare a more general model to
analogous efforts in the literature. This comparison
provides further insight into the usefulness
of the multi-fluid framework and the interpretation of the various coefficients.

So far we have neglected the viscous stresses in the entropy momentum equation. Let us now relax this assumption, i.e.,
account also for $D^\s_{ij}$. In order to simplify the analysis somewhat, we will still assume that $\mathcal{S}^{\n\n} = \sigma^{\n\n}=0$.
That is, we ignore the coupling to vorticity.

We now need
\be
D^\s_{ij} = D_{ij} - D^\n_{ij}  = - g_{ij} [ (\zeta - \zeta^\n) \Theta_\s +
(\zeta^\n - \zeta^{\n\s}) \Theta_{\n\s} ] - 2 [  ( \eta - \eta^\n) \Theta^\s_{ij} +
(\eta^\n - \eta^{\n\n}) \Theta^{\n\s}_{ij} ] \ .
\ee
Rewriting this expression in terms of the particle velocity $v_\n^i$ and the relative flow $w_{\n\s}^i$ we have
\be
D^\s_{ij} = - g_{ij} (\zeta - \zeta^\n) \Theta_\n  - 2 (\eta - \eta^\n) \Theta^\n_{ij}
- g_{ij} \bar{\zeta} \Theta_{\n\s} - 2 \bar{\eta} \Theta^{\n\s}_{ij} \ ,
\label{ds}\ee
where we have introduced
\be
\bar{\zeta} = 2 \zeta^\n - \zeta^{\n\s} - \zeta \ , \qquad \mbox{ and } \qquad \bar{\eta} = 2 \eta^\n - \eta^{\n\s} - \eta \ .
\ee

Again focussing on the case of a rigid solid, the first two terms in \eqref{ds} vanish and we also have
\be
\Theta_{\n\s} = - \nabla_i  w^i \ ,
\ee
and
\be
\Theta^{\n\s}_{ij} =  { 1 \over 2} \left[ \nabla_i w_j + \nabla_j w_i - { 2 \over 3} \nabla_l w^l \right] \ ,
\ee
Adding the relevant terms to (\ref{cat1}), taking $\bar{\zeta}$ and $\bar{\eta}$ to be constant, for simplicity, we have
\begin{multline}
- \partial_t (2 \alpha w_i ) - \nabla_j \left( 2 \alpha w^j w_i \right)
-  \alpha   \nabla_i  w^2  + s \nabla_i T  \\
+   2  \mathcal{R} w_i + \bar{\eta} \nabla^2  w_i + \left( \bar{\zeta} + { 1 \over 3} \bar{\eta} \right) \nabla_i \nabla_j  w^j
 = 0 \ .
\label{cat2}
\end{multline}
We can simplify things further by i) assuming that all parameters are constant, and ii) linearising in $q^i$ (the
caveats regarding these assumptions remain as before). This leads to the equation
\be
- {2 \alpha \over sT} \partial_t q_i + s \nabla_i T
+   {2  \mathcal{R} \over sT} q_i + {\bar{\eta} \over sT} \nabla^2 q_i + { 1 \over sT} \left( \bar{\zeta} + { 1 \over 3} \bar{\eta} \right)  \nabla_i \nabla_j  q^j
 = 0 \ ,
\ee
or, making use of the thermal relaxation time and the thermal conductivity,
\be
\tau \partial_t q_i + \kappa \nabla_i T
+  q_i + l_\eta^2 \nabla^2 q_i + \left( l_\zeta^2 + { 1 \over 3} l_\eta^2 \right)  \nabla_i \nabla_j  q^j
 = 0 \ .
\label{phoneq}\ee
We have also introduced the two lengthscales
\be
l_\eta^2 = {\bar{\eta} \over 2 \mathcal{R}} \ ,
\ee
and
\be
l_\zeta^2 = {\bar{\zeta} \over 2 \mathcal{R}} \ .
\ee
These can be taken to represent the mean-free path associated with the dissipative terms.

This result can be directly compared to the ``phonon hydrodynamics'' model developed by \citet{gk}
(see \citet{llebot}, and \citet{cimmelli} for alternative descriptions). Their model is the most celebrated attempt to account for non-local
heat conduction effects. It accounts for interaction of phonons with each other and the lattice.
Resistive terms are represented by $\tau$ while momentum conserving interactions are associated with
$l_\eta$ and $l_\zeta$. Our model, \eqref{phoneq}, completely reproduces
the \cite{gk} result, once we set
\be
l_\zeta^2 = { 1 \over 3} l_\eta^2 \ .
\ee
This leads to
\be
\tau \partial_t q_i + \kappa \nabla_i T
+  q_i + l_\eta^2  \left( \nabla^2 q_i +2  \nabla_i \nabla_j  q^j \right)
 = 0 \ .
\ee
The usefulness of this result is due to the fact that it can be used both in the
collision dominated and the ballistic phonon regime. In the former, the resistivity dominates, the nonlocal terms can be neglected and heat propagates as waves. In the opposite regime, the momentum conserving interactions
are dominant and we can neglect the thermal relaxation. In this regime, heat propagates by diffusion.
The transition between these two extremes has recently been discussed by \citet{vasquez}.

Interestingly, the non-local heat conduction model may be useful in the
description of nano-size systems. If a system has characteristic size $L$, and $l_\eta/L\gg1$, then one would not necessarily expect a fluid model to apply.
Nevertheless, \citet{alvarez09} have argued that the expected behaviour of the thermal conductivity
as the size of the system decreases (as discussed by \citet{alvarez07} one would expect the ``effective''
conductivity to scale as
$L/l_\eta$) can be reproduced from \eqref{phoneq} provided that the appropriate
slip condition for $q^i$ is applied at the boundaries. A key part of this analysis is the close analogy between
\eqref{phoneq} and the Navier-Stokes equation. In the latter case it is well-known that an applied non-slip
condition at a surface leads to the formation of a viscous boundary layer that dominates the dissipation of
the bulk flow. It appears that the heat problem is quite similar in the ballistic phonon regime,
although the required slip condition is different. This is an interesting problem that requires more detailed study.

\section{The general two-component model}

At this point we have demonstrated that the multi-fluid formalism, with one fluid representing
the massless entropy, reproduces a number of non-trivial results for heat conductivity in rigid solids. However, this is
a simplified problem since one of the degrees of freedom in the system was ``frozen''. In order to complete the
model, we will now relax this assumption and allow $v_\n^i \neq 0$. This leads to a system of equations
with interesting applications. In particular, one could imagine modelling systems with spatial
transitions to superfluidity (as in a neutron star core). In one regime, the system would be dominated by
resistivity, while the thermal relaxation timescale determines the dynamics elsewhere. A unified
model for this problem could prove very useful indeed.

Let us return to the general  two-fluid model and focus on the matter degree of freedom.
As in the case of superfluid Helium \citep{helium}, it is natural to work with the total momentum equation.
By combining \eqref{eulern} and \eqref{eulers} we have
\begin{multline}
\partial_t (\pi_i^\n + \pi_i^\s) + \nabla_l ( v_\n^l \pi_i^\n + v_\s^l \pi_i^\s) + n\nabla_i \mu_\n
+ s \nabla_i T  \\
- n \nabla_i \left( {1 \over 2} m v_\n^2 \right) + \pi_l^\n \nabla_i v_\n^l + \pi_l^\s \nabla_i v_\s^l =
-\nabla_j D^{j}_{\ i} \ .
\end{multline}
We can rewrite this using
\be
\pi_i^\n + \pi_i^\s = \rho v^\n_i \ ,
\ee
\be
 \pi_l^\n \nabla_i v_\n^l + \pi_l^\s \nabla_i v_\s^l = n \nabla_i \left( {1 \over 2} m v_\n^2 \right) - 2 \alpha w_l^{\n\s} \nabla_i w_{\n\s}^l \ ,
\ee
and
\be
 v_\n^l \pi_i^\n + v_\s^l \pi_i^\s = mnv_\n^l v_i^\n - 2\alpha w_{\n\s}^l w_i^{\n\s} \ .
\ee
We also use the fact that the pressure follows from \citep{helium}
\be
\nabla_i p =  n\nabla_i \mu_\n + s \nabla_i T - \alpha \nabla_i w_{\n\s}^2 \ .
\ee
Combining these results we arrive at
\be
\partial_t (\rho v_i^\n) + \nabla_l (\rho v_\n^l v^\n_i)
- \nabla_l ( 2\alpha w_{\n\s}^l w_i^{\n\s}) + \nabla_i p  = -\nabla_j D^{j}_{\ i} \ .
\ee
Using our definition for the heat flux, \eqref{wdef}, we see that if we linearise in $w_i$
(or equivalently, $q_i$) then the problem simplifies considerably. If we also use the continuity equation (\ref{conn}),  we arrive at
an equation that resembles the standard Navier-Stokes result \citep{landau};
\be
\rho(\partial_t + v_\n^l \nabla_l ) v^\n_i + \nabla_i p = -\nabla_j D^{j}_{\ i} \ .
\label{euler}\ee
The right-hand side is, however, different. Keeping $\mathcal{S}^{\n\n} = \sigma^{\n\n}=0$ (as before)
we find that
\be
 -\nabla_j D^{j}_{\ i} = - \left( \tilde{\zeta} + { 1 \over 3} \tilde{\eta} \right) \nabla_i (\nabla_j w_{\n\s}^j) - \tilde{\eta}\nabla^2 w^{\n\s}_i
+ \left( \zeta + { 1 \over 3} \eta \right) \nabla_i (\nabla_j v_\n^j) + \eta\nabla^2 v^\n_i \ ,
\ee
where we have defined
\be
\tilde{\zeta} = \zeta^\n - \zeta \ , \qquad \mbox{ and} \qquad \tilde{\eta} = \eta^\n - \eta \ .
\ee
For simplicity, we have also assumed that the equilibrium configuration is uniform. This leads to
the final result
\begin{multline}
\rho(\partial_t + v_\n^l \nabla_l ) v^\n_i + \nabla_i p \\
=  \left( \zeta + { 1 \over 3} \eta \right) \nabla_i (\nabla_j v_\n^j) + \eta\nabla^2 v^\n_i
- { 1 \over sT} \left( \tilde{\zeta} + { 1 \over 3} \tilde{\eta} \right) \nabla_i (\nabla_j q^j)
- {\tilde{\eta} \over sT} \nabla^2 q_i \ .
\label{mom1}\end{multline}
The first two terms on the right-hand side are familiar from the Navier-Stokes equation \citep{landau}, but
the last two terms are new. They represent the dissipative coupling between the total momentum
and the heat flux. An interesting question concerns whether there are situations where
these terms have decisive impact on the dynamics. Are there, for example, situations where
the last term is similar in magnitude to the second term?

To complete the model, we need the momentum equation for the heat flux. Starting from \eqref{masters}, linearising in the heat flux and
using the results from the previous section, we find that
\begin{multline}
( \partial_t + v_\n^j \nabla_j ) q_i + { 1 \over \tau} q_i + q_i \nabla_j v_\n^j + q_j \nabla_i v_\n^j \\ =
- { \kappa \over \tau} \nabla_i T - { sT \over 2 \alpha} \left[
\left( \tilde{\zeta} + { 1 \over 3} \tilde{\eta} \right) \nabla_i (\nabla_j v_\n^j) + \tilde{\eta}\nabla^2 v^\n_i \right] \\
+ { 1 \over 2 \alpha} \left[\left( \bar{\zeta} + { 1 \over 3} \bar{\eta} \right) \nabla_i (\nabla_j q^j) +  \bar{\eta}  \nabla^2 q_i \right] \ .
\label{mom2}\end{multline}
Comparing to the corresponding equation for a rigid solid, we recognize several terms. Some dissipative terms are, however, new.

Before we conclude our discussion, it is worth making the following observation. In our formulation of the problem we have assumed that the
entropy conservation law is used explicitly. An alternative, more common, strategy is to use the energy equation.
The two approaches are, in principle, equivalent. Nevertheless, it is useful to complement our analysis with
a brief consideration of the energy equation.
From \citet{monster} we have (for an isolated system) the energy equation
\be
\partial_t \mathcal{U} + \nabla_i Q^i = 0 \ .
\ee
After some work, using the various definitions, we find that
\be
Q^i = \left( { 1 \over 2} \rho v_\n^2 - 2 \alpha w_{\n\s}^2 + n \mu + sT \right) v_\n^i - \left( sT + 2\alpha v_\s^j w^{\n\s}_j \right) w_{\n\s}^i \ .
\ee
Here, we can use the fundamental relation
\be
p + E = n\mu + sT \ ,
\ee
to get
\be
Q^i = \left( { 1 \over 2} \rho v_\n^2 - 2 \alpha w_{\n\s}^2 + p + E\right) v_\n^i - \left( sT + 2\alpha v_\s^j w^{\n\s}_j \right) w_{\n\s}^i \ .
\ee
We also have
\be
\mathcal{U} = { 1 \over 2} \rho v_\n^2 - 2\alpha w_{\n\s}^2 + E \ .
\ee
Combining these results, and linearising in $w_{\n\s}^i$, we immediately arrive at the standard energy equation (c.f. \citet{landau})
\be
\partial_t \left( { 1\over 2} \rho v_\n^2 + E \right) + \nabla_i \left[ \left({ 1\over 2} \rho v_\n^2 p+E \right) v_\n^i \right] =  \nabla_i \left( sT w_{\n\s}^i \right) = - \nabla_i q^i \ .
\ee
This confirms, at least at the linear level, the definition \eqref{qdef} of the heat flux.

We now have a ``complete'' model for a heat conducting fluid. It combines the  equations of motion (\ref{mom1}) and
(\ref{mom2}) with the two continuity equations (\ref{conn}) and (\ref{cons}). This model should be relevant
for dynamics on timescales such that the thermal relaxation cannot be ignored. In a typical system, this
would correspond to the extreme high-frequency regime. However, as we already know from the discussion of \citet{helium} the model also applies to superfluid condensates at finite temperatures. In essence, we have a unified
framework for modelling the transition to superfluidity. Finally, there may be situations where the model
applies even if the thermal relaxation can be safely ignored. This would be the case when the ``phonon''
mean free path exceeds the size of the system. As discussed in the previous section, the dissipative terms
in \eqref{mom2} then play the leading role and the crucial boundary effects may be incorporated by imposing
suitable surface conditions on the heat flux. This possibility has so far been discussed only for nano-systems,
but it is worth noting that it may be relevant also for large scale systems. In fact, the ballistic phonon
regime should apply to cold superfluid condensates in neutron stars. To what extent the present analysis
can be applied to that problem is an interesting question for the future.

\section{Concluding remarks}

We have discussed heat conductivity from the point of
view of the variational
multi-fluid model developed by \citet{prix} and \citet{monster}.
We have shown that a two-fluid model that distinguishes between a
massive fluid component and a massless entropy can reproduce a
number of key results from extended irreversible thermodynamics.
In particular, we have demonstrated that the entropy entrainment, that played a central role in our recent discussion of superfluid Helium \citep{helium},
is intimately
linked to the thermal relaxation time that is required to make heat propagation in solids causal. We have also considered non-local terms that arise naturally in the dissipative
multi-fluid model, and related them to models of
phonon hydrodynamics. This discussion may provide useful insight into
the modelling of both nano-systems and superfluids at low
temperatures where the phonon mean-free path is large compared
to the size of the system. Finally, we formulated a ``complete''
heat conducting
two-fluid model and identified a number of ``new'' dissipative terms. Future work
needs to establish whether there are physical situations where these terms play a decisive role.

What is the importance of this work? First of all, we believe that
the discussion provides strong support for the main assumptions of our model: That one can
treat the entropy as an additional fluid,
endowed with the inertial and dynamical properties generally associated with fluids. The entrainment
between particles and the massless entropy plays a key role in this approach.
The connection between this entropy entrainment
and the thermal relaxation time provides an immediate interpretation, and illustrates the importance, of the main  parameter of the model. The
simple fact that the \underline{same} mathematical framework can be used to
model both heat conduction and finite temperature superfluids \citep{helium}
is, in our view, clear evidence of the elegance and promise of the variational
multi-fluid approach. Moreover, since our model has its origin in a fully
relativistic variational analysis, see \citet{livrev} for a review,
the present discussion suggests a promising strategy for developing a
causal relativistic model for heat conductivity. This is known to
be a challenging problem where a number of issues remain to be resolved.

\begin{acknowledgements}
NA acknowledges support from STFC via grant number PP/E001025/1. GLC acknowledges partial support from NSF via
grant number PHYS-0855558.
\end{acknowledgements}


\begin{thebibliography}{25}

\bibitem[{{Alvarez \& Jou}(2007)}]{alvarez07}
Alvarez, F. X., \& Jou, D.,  2007, Memory and nonlocal effects in heat transport: From diffusive to ballistic regimes  , Appl. Phys. Lett. {\bf 90} 083109

\bibitem[{{Alvarez et al}(2008)}]{alvarez}
Alvarez, F. X.,  Casas-V\'azquez, J.,  \&  Jou, D., 2008, Robustness of the nonequilibrium entropy related to the Maxwell-Cattaneo heat equation, Phys. Rev. E {\bf 77} 031110

\bibitem[{{Alvarez et al}(2009)}]{alvarez09}
Alvarez, F. X., Jou, D., Sellitto, A., 2009, Phonon hydrodynamics and phonon-boundary scattering in nanosystems, J. Appl. Phys. {\bf 105} 014317

\bibitem[{{Andersson \& Comer}(2001)}]{ac01}
Andersson, N., \& Comer, G. L., 2001, On the dynamics of superfluid neutron star cores,  MNRAS {\bf 328},
1129

\bibitem[{{Andersson \& Comer}(2006)}]{monster}
Andersson, N., \& Comer, G. L., 2006, A flux-conservative formalism for convective and dissipative multi-fluid systems, with application to Newtonian superfluid neutron stars, Class. Quantum Grav. {\bf 23} 5505

\bibitem[{{Andersson \& Comer}(2007)}]{livrev}
Andersson, N., \& Comer, G. L., 2007, Relativistic Fluid Dynamics: Physics for Many Different Scales, Living Rev. Relativity {\bf 10}

\bibitem[{{Andersson \& Comer}(2009)}]{helium}
Andersson, N., \& Comer, G. L., 2009, {\em Entropy entrainment and dissipation in finite temperature superfluids} in preparation

\bibitem[{{Carter}(1989)}]{carter}
 Carter, B., 1989, ``Covariant Theory of Conductivity in Ideal Fluid or Solid Media'',
  in A. Anile and M. Choquet-Bruhat, eds., {\em Relativistic Fluid Dynamics
  (Noto, 1987)},  pp. 1--64, Heidelberg: Springer-Verlag

\bibitem[{{Casas-V\'azquez \& Jou}(2003)}]{cvj}
Casas-V\'azquez, J., \& Jou, D., 2003, Temperature in non-equilibrium states: a review of open problems and current proposals, Rep. Prog. Phys. {\bf 66} 1937

\bibitem[{{Cattaneo}(1948)}]{cattaneo}
Cattaneo, C., 1948, Atti Seminario Univ. Modena {\bf 3} 33

\bibitem[{{Cimmelli}(2007)}]{cimmelli}
Cimmelli, V. A., 2007, An extension of Liu procedure in weakly nonlocal thermodynamics, J. Math. Phys. {\bf 48} 113510

\bibitem[{{Drew \& Passman}(2007)}]{drew}
Drew, D. A., \& Passman, S. L., 1998, {\em Theory of multicomponent fluids} (Springer, Berlin)

\bibitem[{{Glampedakis \& Andersson}(2009)}]{glitch}
Glampedakis, K., \&  Andersson, N., 2009, Hydrodynamical Trigger Mechanism for Pulsar Glitches, Phys. Rev. Lett. {\bf 102} 141101

\bibitem[{{Glampedakis et al}(2007)}]{prec1}
Glampedakis, K., Andersson, N., \& Jones, D.I., 2007, Stability of Precessing Superfluid Neutron Stars, Phys. Rev. Lett {\bf 100} 081101

\bibitem[{{Guyer \& Krumhansl}(1966)}]{gk}
Guyer, R. A., \&  Krumhansl, J. A., 1966, Thermal Conductivity, Second Sound, and Phonon Hydrodynamic Phenomena in Nonmetallic Crystals
, Phys. Rev. {\bf 148} 778

\bibitem[{{Haskell et al}(2009)}]{erratum}
Haskell, B., Andersson, N., \& Comer, G. L.,  2009, in preparation

\bibitem[{{Jou \& Casas-V\'azquez}(1988)}]{jcv}
Jou, D.,\& Casas-V\'azquez, J., 1988, Extended irreversible thermodynamics of heat conduction, Eur. J. Phys. {\bf 9} 329

\bibitem[{{Jou et al}(1993)}]{joubook}
Jou, D., Casas-V\'azquez, J., \& Lebon, G., 1993, {\em Extended irreversible thermodynamics} (Springer, Berlin)

\bibitem[{{Jou et al}(2004)}]{jlmp}
Jou, D., Lebon, G., Mongiovi, M. S.,  \&  Peruzza, R. A., 2004, Entropy flux in non-equilibrium thermodynamics, Physica A {\bf 338} 445

\bibitem[{{Khalatnikov}(1965)}]{khalatnikov}
Khalatnikov, I. M., 1965, {\em An introduction to the theory of superfluidity}

\bibitem[{{Landau \& Lifshitz}(1959)}]{landau}
Landau, L. D., \& Lifshitz, E. M., 1959, {\em Fluid mechanics} (Oxford, Butterworth Heinemann)

\bibitem[{{Lebon et al}(2008a)}]{lebonbook}
Lebon, G., Jou, D., \& Casas-V\'azquez, J., 2008a, {\em Understanding non-equilibrium thermodynamics} (Springer, Berlin)

\bibitem[{{Lebon et al}(2008b)}]{lrv}
Lebon, G., Ruggieri, M., \& Valenti, A., 2008b, Extended thermodynamics revisited: renormalized flux variables and second sound in rigid solids, J. Phys. Condens. Matter {\bf 20} 025223

\bibitem[{{Llebot et al}(1983)}]{llebot}
Llebot, J.E., Jou, D., \&  Casas-V\'azquez, J., 1983, A thermodynamic approach to heat and electric conduction in solids, Physica A, {\bf 121} 552 (1983)

\bibitem[{{Morro \& Ruggeri}(1987)}]{mr87}
Morro, A., \& Ruggeri, T., 1987, Int. J. Non-Linear Mechanics, {\bf 22} 27

\bibitem[{{Morro \& Ruggeri}(1988)}]{morro}
Morro, A., \& Ruggeri, T., 1988, J. Phys. C, Non-equilibrium properties of solids obtained from second-sound measurements, {\bf 21} 1743

\bibitem[{{M\"uller \& Ruggeri}(1993)}]{mullerbook}
M\"uller, I., \& Ruggeri, T., 1993, {\em Extended thermodynamics}, (Springer, New York)

\bibitem[{{Passamonti et al}(2009)}]{andrea}
Passamonti, A., Haskell, B., and Andersson, N., 2009, Oscillations of rapidly rotating superfluid stars, MNRAS {\bf 396} 951

\bibitem[{{Prix}(2004)}]{prix}
Prix, R., 2004, Variational description of multifluid hydrodynamics: Uncharged fluids, Phys. Rev. D {\bf 69} 043001

\bibitem[{{Rajagopal \& Tao}(1995)}]{raja}
Rajagopal, K. R., \& Tao, L., 1995, {\em Mechanics of mixtures} (World Scientific, Singapore)

\bibitem[{{Ruggeri et al}(1996)}]{rms}
Ruggeri, T., Muracchini, A., \& Seccia, L., 1996,  Second sound and characteristic temperature in solids, Phys. Rev. B, {\bf 54} 332

\bibitem[{{V\'asquez \& M\'arkus}(2009)}]{vasquez}
V\'asquez, F. \& M\'arkus, F., 2009, Size effects on heat transport in small systems: Dynamical phase transition from diffusive to ballistic regime, J. Appl. Phys. {\bf 105} 064915


\end{thebibliography}
\end{document}